# Condensés de textes par des méthodes numériques

Juan-Manuel Torres[1,2,3], Patricia Velázquez-Morales[2], Jean-Guy Meunier[3]

[1] École Polytechnique/DGI – CP 6079 Succ. Centre-ville – H3C3A7 Montréal – Canada
[2] ERMETIS/Univ. du Québec – 555 Boul. de l'Université – G7H2B1 Chicoutimi – Canada
[3] LANCI/Univ. du Québec – CP 8888 Succ. Centre-Ville – H3C3P8 Montréal – Canada

## Abstract

Since information in electronic form is already a standard, and that the variety and the quantity of information become increasingly large, the methods of summarizing or automatic condensation of texts is a critical phase of the analysis of texts. This article describes Cortex a system based on numerical methods, which allows obtaining a condensation of a text, which is independent of the topic and of the length of the text. The structure of the system enables it to find the abstracts in French or Spanish in very short times.

## Résumé

Étant donné que la variété et la quantité de l'information sous forme électronique deviennent de plus en plus grandes, des méthodes d'obtention de résumés ou de condensation automatique de textes constituent une phase critique de l'analyse de textes. Cet article décrit Cortex, un système basé sur des méthodes numériques qui permet l'obtention d'un condensé d'un texte, qui est indépendant du thème, de l'ampleur du texte et de la façon dont il est écrit. La structure du système lui permet de trouver la condensation de textes multilangues dans des temps très courts. Des applications en français ou espagnol sont présentées et analysées.

**Keywords:** Condensés de textes, résumés automatiques, analyse de textes, catégorisation, méthodes statistiques.

## 1. Introduction

L'information textuelle électronique s'accumule rapidement et en très grande quantité. Alors les documents sont catégorisés d'une façon très sommaire. Le manque de standards est un facteur critique, et tous les analyses des textes (dépistage, exploration, récupération, résumés, etc.) sont des taches extrêmement difficiles (Torres-Moreno et al., 2000). C'est pourquoi des méthodes d'obtention de résumés automatique de textes constituent une phase cruciale de l'analyse de textes. Les méthodes linguistiques sont pertinentes dans ces tâches, mais leur utilisation concrète demeure encore difficile (en raison de plusieurs facteurs comme l'ampleur, la dynamique des corpus) ou limitée à des domaines restreints (Saggion and Lapalme, 2000). D'un autre côté, des méthodes statistique-neuronales sont de plus en plus utilisées dans plusieurs domaines du traitement de l'information textuelle (Salton, 1971; Salton and McGill, 1983; Deerwester et al., 1990; Leloup, 1997; Veronis et al., 1991; Balpe et al., 1996; Torres-Moreno et al., 2000; Memmi et al., 1998; Meunier and Nault, 1997; Memmi and Meunier, 2000). Cet article présente un étude basé sur l'approche vectorielle des textes (Salton and McGill, 1983) pour obtenir des condensés pertinents de documents. La forme la plus connue et la plus visible de la condensation de textes est le résumé, représentation abrégée et exacte du contenu d'un document (ANSI, 1979). Étant donné que l'état de l'art ne permet d'obtenir que de résumés informatifs (Morris et al., 1999), nos recherches porteront sur l'obtention de ce type de condensés. Nous allons présenter un algorithme récemment développé. Il s'agit d'une chaîne de traitement numérique qui combine plusieurs traitements statistiques et informationnels (comme des calculs d'entropie, le poids fréquentiel des segments et des mots, et plusieurs mesures d'Hamming parmi d'autres) avec un algorithme optimal de décision pour choisir des phrases pertinents du texte. L'ensemble de ces phrases rendent ce qu'on appelle le condensé du document.

## 2. Pré-traitement

Dans l'approche vectorielle on traite de textes dans leur ensemble, en passant par une représentation numérique très différente d'une analyse structurale linguistique, mais qui permet des traitements performants (Memmi, 2000). L'idée consiste à représenter les textes dans un espace approprié et à leur appliquer des traitements vectoriels. La chaîne **Conterm** (Seffah and Meunier, 1996) comporte un ensemble des processus de filtrage, segmentation et lemmatisation. Par opposition à l'analyse symbolique classique, ces processus sont très performants et peuvent être appliqués à de gros corpus. Nous avons adapté les processus de **Conterm** a nos besoins, ainsi un module de pre-traitement a été développé (Gravel et al., 2001). Le texte original comporte $N_M$ mots (mots fonctionnels, de noms ou de verbes fléchis et des mots composés). On emploie la notion de **terme** pour designer un mot plus abstrait (Memmi, 2000). Pour réduire la complexité des processus de réduction de filtrage du lexique sont amorcés[1]. La lemmatisation de verbes des langues à morphologie variable (langues romances) s'avère très important pour la réduction du lexique, et consiste à trouver la racine des verbes fléchis et à ramener les mots au singulier masculin avant de les compter[2]. Ce processus permet de diminuer la malédiction dimensionnelle qui pose de sérieux problèmes dans des grandes dimensions. Nous explorons aussi d'autres voies pour la réduction du lexique en utilisant des processus de synonimisation à l'aide de dictionnaires. La segmentation est faite en utilisant des séparateurs (<!>,<?>,<.>,<:>)[3]. Un indice de repérage importante d'information est le titre d'un document. Toutefois nous expériences ont été réalisées sur de textes bruts, donc les titres, sous-titres et sections ne sont pas marqués explicitement (comme ce le cas des formats HTML ou XML). Après le pré-traitement, le nouveau texte comporte $P$ segments avec $N_f$ termes totaux.

## 3. Condensation du texte

La segmentation transforme un document dans un ensemble de vecteurs $\vec{\Xi} = (\Xi_1, \Xi_2, \cdots, \Xi_{N_M})$ $= \{0, 1\}^{N_M}$. Chaque segment de texte est représenté par un vecteur à composantes binaires. La dimension $N_M$ est le nombre total de mots différents. L'ensemble des segments dont on dispose consiste en $P$ vecteurs et la matrice $\Xi = \vec{\Xi}^\mu; \mu = 1, \cdots, P$ représente le texte. Seulement les termes à fréquence supérieure à 1 ont été utilisés (Torres-Moreno et al., 2001), et donc un lexique de $N_\mathcal{L}$ termes est obtenu. On garde la relation $N_\mathcal{L} \leq N_f \leq N_M$. Nous avons donc défini :

$$\rho_\mathcal{L} = \frac{N_\mathcal{L}}{N_M} \qquad (1)$$

---

[1] La suppression des mots fonctionnels, des mots à haute et très basse fréquence d'apparition, suivi par la suppression du texte entre parenthèses, de chiffres et des symboles.

[2] Ainsi on pourra ramener à la même forme **chanter** les mots *chantaient, chanté, chanteront* et eventuellement *chante* et *chanteur*

[3] Le critère de segments à taille fixée a été écarté, car on cherchait l'extraction des phrases complètes.

comme le ratio de réduction du lexique filtré/lemmatisé. La matrice Terme-Segment $\xi = \vec{\xi}^\mu; \mu = 1, \cdots, P$, dérivée de $\Xi$ représente le lexique réduit du texte :

$$\xi = \begin{bmatrix} \xi_1^1 & \xi_2^1 & \xi_3^1 & \cdots & \xi_{N_\mathcal{L}}^1 \\ \xi_1^2 & \xi_2^2 & \xi_3^2 & \cdots & \xi_{N_\mathcal{L}}^2 \\ \vdots & \vdots & \vdots & & \vdots \\ \xi_1^\mu & \xi_2^\mu & \xi_3^\mu & \cdots & \xi_{N_\mathcal{L}}^\mu \\ \vdots & \vdots & \vdots & \ddots & \vdots \\ \xi_1^P & \xi_2^P & \xi_3^P & \cdots & \xi_{N_\mathcal{L}}^P \end{bmatrix} \qquad (2)$$

Dans cette matrice chaque composante montre la présence ($\xi_i^\mu = 1$) ou l'absence ($\xi_i^\mu = 0$) du mot $i$ dans un segment $\mu$. De façon analogue, la matrice fréquentielle :

$$\gamma = \vec{\gamma}^\mu; \mu = 1, \cdots, P \qquad (3)$$

où chaque composante $\vec{\gamma} = (\gamma_1, \gamma_2, \cdots, \gamma_{N_\mathcal{L}})$ contient la fréquence $\gamma_i^\mu$ du terme $i$ dans un segment $\mu$. Cette matrice contient l'information fréquentielle essentielle du texte. La condensation du texte va s'effectuer sur ces deux matrices, qui constituent l'espace des entrées au système. Nous avons défini la **taille réduite** des matrices $\gamma$ et $\xi$ comme :

$$\alpha = \frac{P}{N_\mathcal{L}} \qquad (4)$$

qui représente la proportion $P$ de segments par rapport à la dimension $N_\mathcal{L}$ du lexique réduit à l'entrée. Les segments possèdent une quantité hétérogène du lexique employé : il y a des segments plus importants que d'autres, qui seront extraits par l'algorithme pour obtenir un condensé.

## 4. Algorithme

La méthode Cortex de résumés automatiques est composée de deux algorithmes : une méthode de construction des métriques informationnelles indépendantes combinée avec un algorithme pour la récupération de l'information codée. Ce dernier prendra une décision sur les segments à choisir en fonction d'une stratégie des votes.

### *4.1. Métriques*

Des informations mathématiques et statistiques importantes sont calculées à partir des matrices $\xi$ (2) et $\gamma$ (3) sous forme de métriques. Elles mesurent la quantité d'information contenue dans chaque segment : plus il est importante, plus il comporte des valeurs élevés.

- Mesures frequentielles.
    - Fréquence des mots. La somme des fréquences des mots par segment calcule un poids spécifique d'un segment $\mu$ en utilisant l'expression :

$$F^\mu = \sum_{i=1}^{N_\mathcal{L}} \gamma_i^\mu \qquad (5)$$

$\gamma_i^\mu$ est la fréquence du mot $i$ dans le segment $\mu$. L'expression :

$$T = \sum_{\mu=1}^{P} \sum_{i=1}^{N_\mathcal{L}} \gamma_i^\mu \qquad (6)$$

corresponde a la taille du lexique fréquentiel contenue $\gamma$. Nous introduisons ici la quantité :

$$\rho_\gamma = \frac{T}{N_M} \qquad (7)$$

définie comme le ratio de réduction du lexique fréquentiel.

– Interaction de segments. Dans chaque segment $\mu$, un mot $i$ qui est présent au même temps dans un ou plusieurs autres segments, on dit qui est en « interaction ». La somme de toutes les interactions de mots de chaque segment constitue alors l'interaction entre segments. Nous la comptabilisons de la façon suivante :

$$I^\mu = \sum_{i=1}^{N_\mathcal{L}} \sum_{\substack{j=1 \\ j \neq \mu}}^{P} \xi_i^j \qquad (8)$$

– Somme fréquentielle des probabilités $\Delta$. Calculons d'abord les probabilités de mots. Soit $p_i$ la probabilité d'apparition du terme $i$ dans le texte :

$$p_i = \frac{1}{T} \sum_{\mu=1}^{P} \gamma_i^\mu \qquad (9)$$

La somme fréquentielle des probabilités est calculé :

$$\Delta = \sum_{i=1}^{N_\mathcal{L}} p_i \gamma_i^\mu \; ; \text{ si } \xi_i^\mu \neq 0 \qquad (10)$$

- Mesures entropiques. L'entropie d'un segment $\mu$ nous la calculons en utilisant :

$$E^\mu = -\sum_{i=1}^{N_\mathcal{L}} x_i^\mu \log_2 x_i^\mu \qquad (11)$$

avec

$$x_i^\mu = \frac{\gamma_i^\mu}{\sum_{i=1}^{N_\mathcal{L}} \gamma_i^\mu} \qquad (12)$$

- Mesures d'Hamming. Une distance de Minskowski a été utilisée comme mesure de base.

  – Les distances d'Hamming $\Psi$. Cette quantité mesure la distance entre paires de mots $i$ et $j$ dans l'espace des segments. Chaque mot étant représentée par un vecteur binaire $\vec{\xi_i} = \{0,1\}^P$. Il faut d'abord, calculer la matrice d'Hamming $H$, qui est une matrice diagonale supérieure à dimension $N$.

$$H_i^{i+1} = \left\{ \begin{array}{l} 1 \; ; \text{ si } \xi_i^\mu \neq \xi_j^\mu \\ 0 \text{ autrement} \end{array} \right\} \begin{array}{l} i=1,\cdots,N_\mathcal{L}-1 \\ j=i+1,\cdots,N_\mathcal{L} \\ \mu=1,\cdots,P \end{array} \qquad (13)$$

Ensuite on calcule la somme des distances d'Hamming :

$$\Psi^\mu = \sum_{i=1}^{N_\mathcal{L}} \sum_{j=i+1}^{N_\mathcal{L}} H_i^j \text{ si } \left(\xi_i^\mu, \xi_j^\mu\right) \neq 0 \qquad (14)$$

  – Le poids d'Hamming des segments. Chaque segment possède un « poids » $\phi^\mu$, qui est égal à la somme des termes présentes dans le segment, c'est-à-dire, dans chaque ligne de la matrice $\xi$ (2) :

$$\phi^\mu = \sum_{i=1}^{N_\mathcal{L}} \xi_i^\mu \; ; \text{ si } \xi_i^\mu \neq 0 \qquad (15)$$

– La somme des poids d'Hamming de mots $\Theta$. De la même manière, on peut mesurer le poids spécifique des mots sur chaque colonne $\mu$ de la matrice $\xi$; $\mu = 1, \cdots, P$, ce qui donne un « Poids d'Hamming des mots » :

$$\psi^i = \sum_{\mu=1}^{P} \xi_i^\mu \text{ ; si } \xi_i^\mu \neq 0 \text{ ; } i = 1, \cdots, N_\mathcal{L} \qquad (16)$$

Ensuite on calcule la somme des poids d'Hamming des mots par segments :

$$\Theta^\mu = \sum_{i=1}^{N_\mathcal{L}} \psi_i \text{ ; si } \xi_i^\mu \neq 0 \qquad (17)$$

- Mesures mixtes. Des mesures de distances combinés avec des mesures frequentielles ont été aussi considérés.
    – Le poids d'Hamming lourd. Il est obtenu de la multiplication du poids d'Hamming du segment $\phi$ par le poids d'Hamming des mots $S_{HM}$ :

$$\Pi^\mu = \phi^\mu S_{HM}^\mu \qquad (18)$$

– Somme des poids d'Hamming de mots par fréquence $\Omega$. Ceci corresponde à la somme des poids d'Hamming des mots $i$ existantes dans chaque segment $\mu$, multiplié par la fréquence correspondante.

$$\Omega^\mu = \sum_{i=1}^{N_\mathcal{L}} \phi^\mu \gamma_i^\mu \text{ ; si } \gamma_i^\mu \neq 0 \qquad (19)$$

*4.2. Algorithme de décision*

Nous avons développé un algorithme pour récupérer l'information codée par les métriques. L'idée est simple : étant donné les votes pour un événement particulier qui provient d'un ensemble de $k$ votants indépendants, chaqu'un avec une certaine probabilité d'avoir raison, trouver la décision optimale. La méthode que nous avons développée s'appelle Algorithme de décision (AD) (Torres-Moreno et al., 2001). L'algorithme de décision utilise deux probabilités mutuellement exclusives : $p_0$ et $p_1$. On présente les $k$ votants en modifiant $p_0$ et $p_1$ en fonction des sorties $\pi_j$; $j = 1, \cdots, k$ sur chaque segment. Toutes les valeurs des métriques ont été normalisées avant d'être utilisées dans l'AD. Cet algorithme de décision possède deux propriétés intéressantes : convergence et amplification [4].

## 5. Expériences et résultats

Nous avons testé notre algorithme sur des articles de vulgarisation scientifique. Les textes extraits de la presse sur Internet sont de petite taille. L'objectif a été d'obtenir un condensé du 25% du nombre de segments. Nous avons comparé avec les logiciels **Minds**[5], **Summarizer**©[6] et avec **Word**©. Dans les cas de **Minds** et **Word**, le paramètre utilisé a été d'obtenir une synthèse du 25% de la taille du texte. Nous avons aussi demandé à 14 personnes de faire un condensé à la main : choisir les phrases du texte qui leur sembleraient les plus pertinentes [7].

---

[4] Les probabilités $p_0$ et $p_1$ sont modifiées en tout temps de façon mutuellement exclusive, et l'écart entre $p_0$ et $p_1$ est changé toujours avec une probabilité $\geq \frac{1}{2}$ de l'améliorer. On amplifié car la probabilité de choisir un segment pertinent est $\geq$ à la probabilité $\pi_j$ du meilleur votant branché à ce moment.

[5] http://messene.nmsu.edu/minds/SummarizerDemoMain.html

[6] http://www.copernic.com

[7] Tous les sujets ont un niveau d'études universitaire et habitués à faire des résumés.

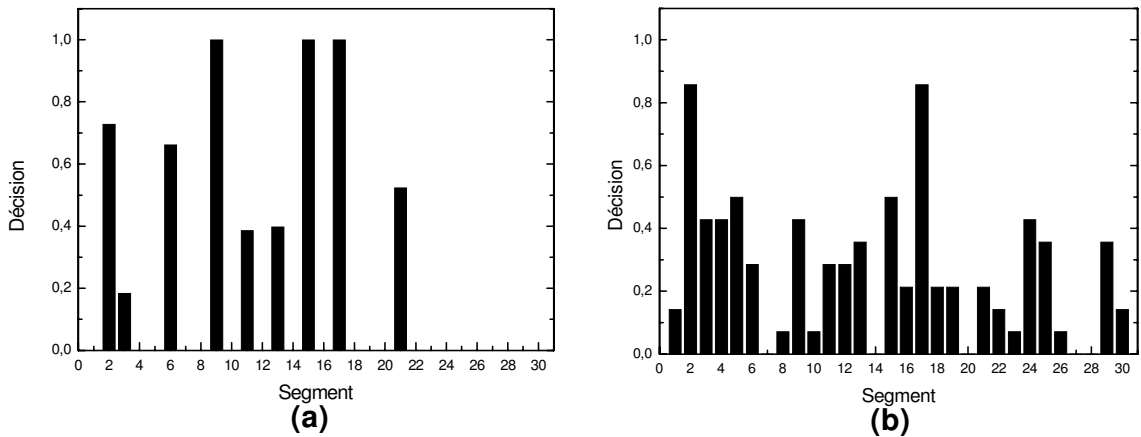

Figure 1: *Choix de segments pertinents pour le texte « Puces » a) par le système* Cortex *: les segments 2 et 17 ont été bien choisis. b) Par les 14 sujets humains.*

### *5.1. Textes en français*

Nous avons étudié le texte « Puces »[8] (artificiellement ambigu et composé d'un mélange non homogène de textes : sujets « puces biologiques » et « puces informatiques ») dans le cadre de la classification de segments par leur contenu (Torres-Moreno et al., 2000). Les segments les plus importantes sélectionnés par les humains sont le 2, 5, 15 et 17 (figure 1b). Nous reproduisons sur la figure (1a) nos résultats, où on voit que les segments importants ont été bien repérés. Cortex montre un résumé équilibré, du même que celui obtenu par **Minds** (même si ce dernier ne trouve ni le segment 5 ni le 15 sur la figure (2a)). Par contre les résultats de **Word** sont biaisés et peu pertinents comme on le voit sur la figure (2b) (Torres-Moreno et al., 2001). Pour le texte « Fêtes »[9] les résultats préliminaires montrent que Cortex trouve de résumés acceptables (figure 3). Nous avons effectué des comparaisons avec **Summarizer** (Huot, 2000), et nos condensés sont comparables voire de meilleure qualité. Dans d'autres tests les condensés trouvés par Cortex semblent être assez cohérentes. Nous avons constaté toujours que les condensés obtenus par les sujets humains dépendent de l'expertise de la personne et des ses capacités d'abstraction, ce qui donne de fois des résultats assez écartés (figure 1b et figure 3b). Malgré cela le choix fait par des humains semble être une référence sur les segments importants, mais notre méthode comparable.

### *5.2. Textes en espagnol*

Nous avons travaillé sur deux textes en espagnol : « Nopal »[10] et « Tabaco »[11]. Les résultats de Cortex sur le texte « Nopal » sont montrés sur les figures 4a et 4b. On constate la bonne qualité du condensé, même dans des textes à très petit lexique. « Nopal » possède seulement $N_{\mathcal{L}} = 5$ mots, $P = 7$ segments et **Word** est incapable de le traiter.

---

[8]http://www.gegi.polymtl.ca/info/jmtomore/pvm/cortex/textos/puces.html
[9]http://www.quebecmicro.com/6-12/6-12-28.html
[10]http://www.invdes.com.mx/suplemento/anteriores/Noviembre2000/htm/espina.html
[11]http://www.invdes.com.mx/suplemento/anteriores/Diciembre2000/htm/tabaco.html

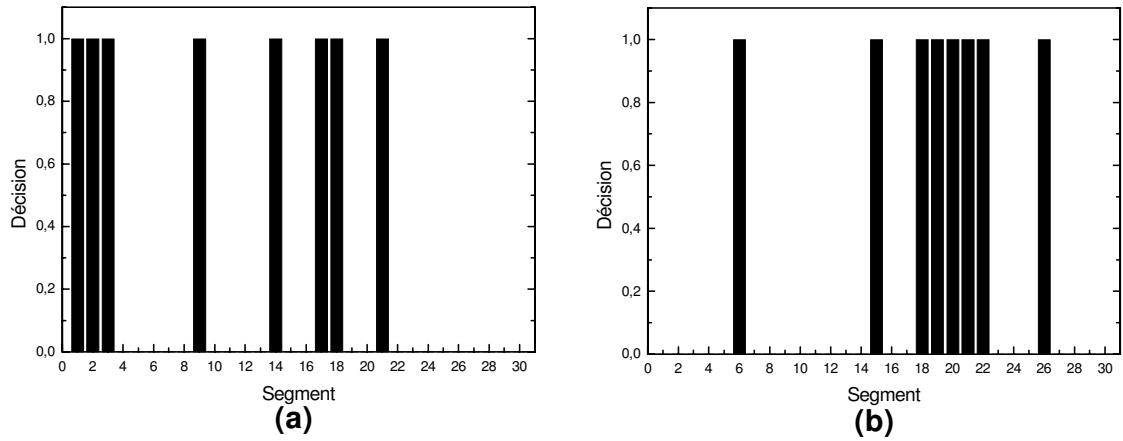

Figure 2: *Choix de segments pour le texte « Puces » a) par le système* **Minds** *et b) par le synthétiseur* **Word**.

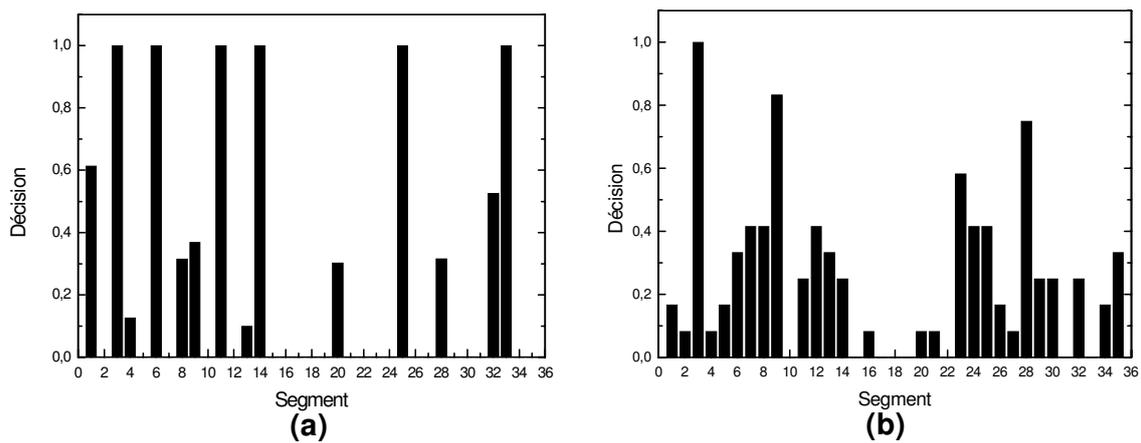

Figure 3: *Choix de segments pertinents pour le texte « Fêtes » a) par le système* Cortex *: plusieurs segments importants ont été choisis. b) Par les 14 sujets humains.*

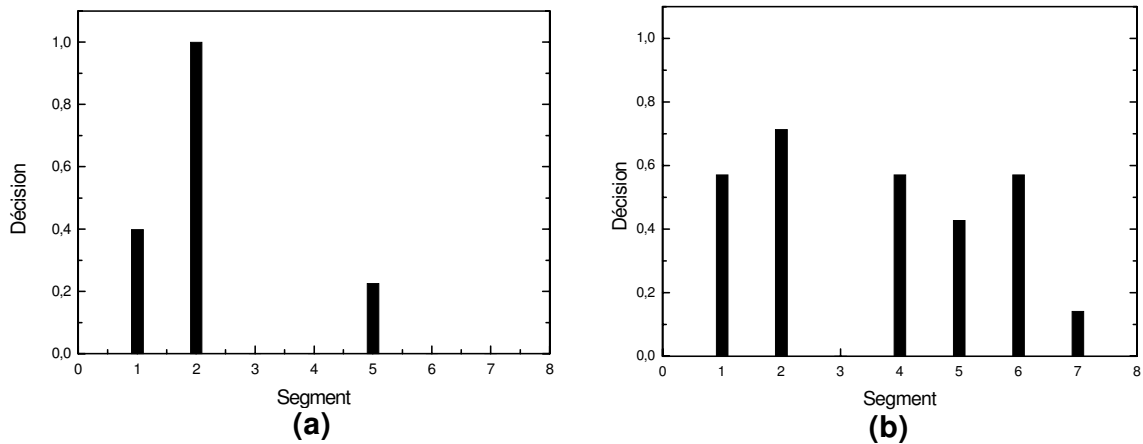

Figure 4: *a) Choix de segments pour « Nopal » fait par* Cortex. *Les segments 1 et 2 qui ont une importance particulière ont été bien repérés. b) Choix fait par 5 sujets humains.*

## 6. Discussion

### 6.1. Taille du lexique

Nous savions que grâce au processus de pre-traitement, le lexique était de plus en plus réduit, c'est-à-dire $N_{\mathcal{L}} \leq N_f \leq N_M$. Des études sur les ratios de réduction moyens du lexique ont été effectués sur l'ensemble de $\tau$ textes (français et espagnol). Ceci nous a permis d'établir expérimentalement des estimateurs $\hat{\rho}_{\mathcal{L}}$ et $\hat{\rho}_\gamma$ pour le lexique filtré/lemmatisé réduit $\rho_{\mathcal{L}}$ (1) et le lexique fréquentiel $\rho_\gamma$ (7) respectivement. Si nous introduisons :

$$\hat{N}_M = \frac{1}{\tau}\sum_{i=1}^{\tau} N_{Mi}; \hat{N}_f = \frac{1}{\tau}\sum_{i=1}^{\tau} N_{fi}; \hat{N}_{\mathcal{L}} = \frac{1}{\tau}\sum_{i=1}^{\tau} N_{\mathcal{L}i}; \hat{T} = \frac{1}{\tau}\sum_{i=1}^{\tau} T_i;$$

Alors :

$$\hat{\rho}_f = \frac{\hat{N}_f}{\hat{N}_M} \; ; \; \hat{\rho}_\gamma = \frac{\hat{T}}{\hat{N}_M} \; ; \; \hat{\rho}_{\mathcal{L}} = \frac{\hat{N}_{\mathcal{L}}}{\hat{N}_M}$$

Nous avons calculé pour les textes en français : $\hat{\rho}_f = 0,414 \pm 0,032$ ; $\hat{\rho}_{\mathcal{L}} = 0,068 \pm 0,015$ et $\hat{\rho}_\gamma = 0,224 \pm 0,058$. Sur les figures (5a) et (5b) nous montrons les valeurs de ces ratios et leurs moyennes sur l'ensemble de textes en français. La réduction de la taille du lexique filtré/lemmatisé $\hat{\rho}_{\mathcal{L}}$ suit un comportement linéaire par rapport au nombre de mots du texte original, donc pour obtenir un condensé d'un texte avec nos méthodes on utilise seulement un seizième du volume de termes totaux du document. Toutes ces réductions permettent de diminuer la malédiction dimensionnelle. Finalement, sur la figure (6) on montre une courbe qui modèle le comportement du lexique essentiel $N_{\mathcal{L}}$ en fonction de la taille originale du texte $N_M$. De façon similaire nous avons calculé la réduction du lexique pour les deux textes en espagnol. Nous avons constaté un comportement semblable à celui des textes en français.

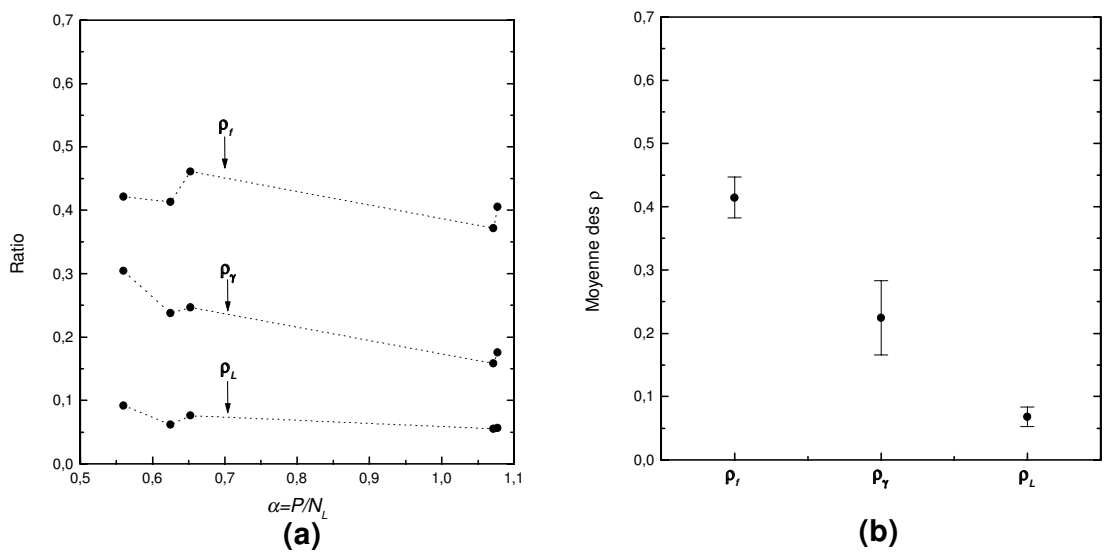

Figure 5: *Textes français. a) Ratios de réduction du lexique en fonction de $\alpha$. $\rho_f$ : réduction après le filtrage/lemmatisation, $\rho_\mathcal{L}$ : réduction de termes de frequence>1, $\rho_\gamma$ : Lexique frequentiel : la fraction de $\gamma$ par rapport au texte original, b) Moyennes des ratios de réduction du lexique français $\rho$, calculés sur l'ensemble de $\tau = 5$ textes.*

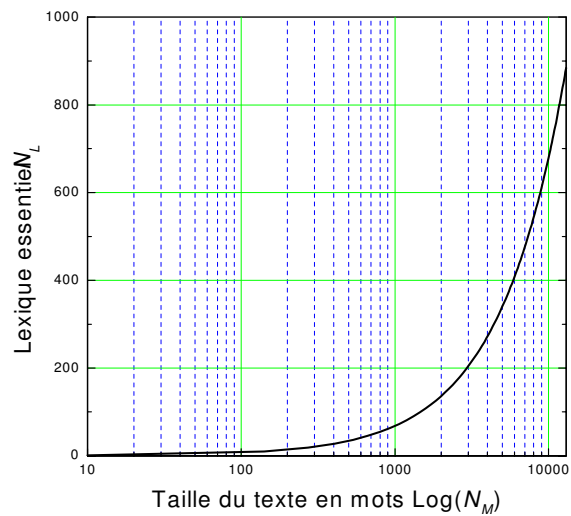

Figure 6: *Lexique essentiel $N_\mathcal{L}$ comme fonction de la taille du texte $N_M$. L'axe horizontale est logarithmique. On observe une croissance parcimonieuse du lexique essentiel.*

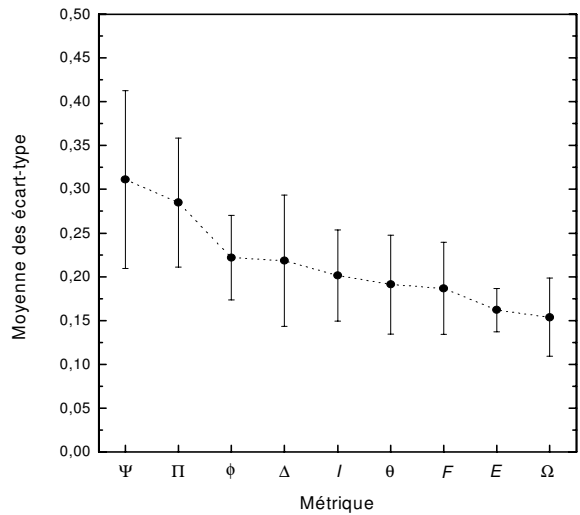

Figure 7: *Moyenne du pouvoir de discrimination des métriques.*

### *6.2. Ordre de présentation des métriques*

Un étude a montré que l'ordre de présentation des métriques a un certain impact sur les performances de l'AD. Initialement l'ordre de présentation de métriques a été arbitraire, puis un étude statistique a été effectué. Leur pouvoir discriminatoire a été mesuré comme fonction des écart-types des métriques par rapport à chaque segment. En effet, il y a des métriques que sont plus discriminantes que d'autres. La somme des distances d'Hamming entre mots à l'intérieur d'un segment est très discriminante. Par contre, les métriques qu'impliquent des calculs d'entropie ou d'une valeur fréquentielle semblent l'être moins. La figure (7) montre les moyennes du pouvoir discriminatoire des métriques. L'ordre de présentation a été : **1.** Les distances d'Hamming $\Psi$, **2.** Les poids d'Hamming lourd $\Pi$, **3.** Le poids d'Hamming des segments $\phi$, **4.** La somme des probabilités par fréquence $\Delta$, **5.** Les interactions $I$, **6.** La somme des poids d'Hamming $\Theta$, **7.** La fréquence $F$, **8.** L'entropie $E$ et **9.** La somme fréquentielle des poids d'Hamming $\Omega$. Nous avons décidé donc d'utiliser cette ordre de présentation des métriques à l'Algorithme de décision, ce qui permet d'obtenir un choix de segments pertinents, stable et cohérent.

### *6.3. Ordre de présentation des segments*

Nos expériences ont montré que l'ordre de présentation des segments n'a aucune influence sur le choix final de l'Algorithme de décision. Nous avons découpé les textes en segments et nous les avons mélangé au hasard pour obtenir un nouveau texte. Ce texte a été présenté à nouveau à Cortex et les mêmes résultats ont été retrouvés aussi bien en français et en espagnol. Par contre, des tests sur **Minds** et **Word** montrent que ces méthodes sont parfois dépendantes de l'ordre de présentation des segments. En effet, la segmentation de phrases par le séparateur < : > a tendance à perturber la pertinence de choix de ces méthodes, mais pas dans le notre.

## 7. Conclusion

L'algorithme Cortex est un condensateur de textes très performant. Cette technologie permet de traiter de vastes corpus, multi-langues (français, espagnol), sans préparation, avec une certaine quantité de bruit, de manière dynamique et en un court lapse de temps. De plusieurs tests faits en comparaison avec des sujets humains ou d'autres méthodes de condensation, ont montré que notre algorithme est capable de retrouver les segments de texte les plus pertinents (indépendante de la taille du texte et des sujets abordés). On obtient ainsi un résumé balancé car la plupart des thèmes sont abordés dans le condensé final. Le logiciel **Summarizer** communique avec l'utilisateur en lui demandant des concepts à retenir dans le résume. Ceci est une approche intéressant qui pourrait être intégré dans notre algorithme de décision basé sur les votes de métriques, et qui est déjà robuste, convergente, amplificateur et indépendant de l'ordre de présentation des segments. Nous pensons que l'ajout d'autres métriques (entropie résiduelle, détection des changements d'entropie, maximum d'entropie) et d'un identificateur automatique de langues, pourraient améliorer la qualité des condensations.

## Remerciements